\documentclass{aa}
\usepackage{amsmath,amssymb}
\usepackage{graphicx}
\usepackage{natbib}
\def\Mpl{M_{\rm P}}

\begin{document}

\title{Non-linear dynamics of the minimal theory of massive gravity}

\author{Robert Hagala$^{1,2}$
\and Antonio De Felice$^3$
\and David F. Mota$^1$
\and Shinji Mukohyama$^{3,4}$}
\institute{$^1$Institute of Theoretical Astrophysics, University of Oslo, PO Box 1029 Blindern, 0315 Oslo, Norway \\ $^2$Expert Analytics AS, Tordenskiolds gate 6, 0160, Oslo, Norway \\
$^3$Center for Gravitational Physics, Yukawa Institute for Theoretical Physics, Kyoto University, 606-8502, Kyoto, Japan \\ $^4$Kavli Institute for the Physics and Mathematics of the Universe (WPI), The University of Tokyo, 277-8583, Chiba, Japan}
\date{\today}
\abstract{We investigate cosmological signatures of the minimal theory of massive gravity (MTMG). To this aim, we simulate the normal branch of the MTMG by employing the \textsc{Ramses} \mbox{$N$-body} code and extending it with an effective gravitational constant $G_{\rm eff}$. We implement an environment-dependent $G_{\rm eff}$ as a function of the graviton mass and the local energy density as predicted by MTMG. We find that halo density profiles are not a good probe for MTMG because deviations from general relativity (GR) are quite small. Similarly, the matter power spectra show deviations only at the percentage level. However, we find a clear difference between MTMG and GR in that voids are denser in MTMG than in GR. As measuring void profiles is quite a complex task from an observational point of view, a better probe of MTMG would be the halo abundances. In this case, MTMG creates a larger amount of massive halos, while there is a suppression in the abundance of small halos.}

\maketitle

{\footnotesize YITP-20-155, IPMU20-0124}

\section{Introduction}

Mass is one of the most fundamental characteristics of particles and fields, and a long-standing problem in classical field theory is whether or not the graviton, the spin-$2$ particle that mediates gravity, can have a non-zero mass. The seminal work by \cite{Fierz:1939ix} uncovered a unique Lorentz-invariant graviton mass term at the level of a linear theory. In 1970, van Dam, Veltman, and Zhakharov~\citep{vanDam:1970vg,Zakharov:1970cc} found a discontinuity at the massless limit of the Fierz-Pauli theory, questioning its viability because the discontinuity would imply an $\mathcal{O}(1)$ deviation from general relativity (GR) however small the mass of the  graviton. While the discontinuity of the massless limit can be resolved by non-linearity as shown in 1972 by Vainshtein~\citep{Vainshtein:1972sx}, Boulware and Deser (BD)~\citep{Boulware:1973my} in the same year pointed out that the same kind of non-linearity that saved the massive gravity from the discontinuity causes a problem, namely that there appears a ghost at the non-linear level. The BD ghost posed a problem for the massive gravity for almost 40 years until de Rham, Gabadadze, and Tolley (dRGT) \citep{deRham:2010ik} in 2010 discovered a fully non-linear classical theory of massive gravity without the BD ghost.

Given a consistent non-linear theory of massive gravity, it is natural to study its implications for cosmology. In particular, as the graviton mass modifies the behaviour of gravity at long distances, it would be interesting to ask whether the modified dynamics can address the mystery of the accelerated expansion of the Universe today. However, it soon turned out that the dRGT theory does not allow an expanding (or contracting) flat Friedmann–Lema\^{i}tre–Robertson–Walker (FLRW) solution~\citep{DAmico:2011eto}. Although an open FLRW solution with self-acceleration was found~\citep{Gumrukcuoglu:2011ew}, it was shown that the solution suffers from ghost instability at the non-linear level~\citep{DeFelice:2012mx}. If we extend the dRGT theory by allowing for a non-Minkowski fiducial metric, then another branch of FLRW solutions can be found. This branch is called the normal branch, in contrast to the previous one called the self-accelerating branch, and suffers from an instability at the linear level called the Higuchi ghost~\citep{Higuchi:1986py,Fasiello:2012rw}. For these reasons, all FLRW solutions in dRGT theory (with an either Minkowski or non-Minkowski fiducial metric) are unstable~\citep{DeFelice:2012mx}.

Progress has been made towards stable cosmological solutions in the framework of non-linear massive gravity. In general, there are two options: to break the symmetry of FLRW (i.e. either homogeneity~\citep{DAmico:2011eto} or isotropy~\citep{Gumrukcuoglu:2012aa,DeFelice:2013awa}) or to extend the theory (see e.g.~\citep{Kenna-Allison:2019tbu} and references therein).

The minimal theory of massive gravity (MTMG)~\citep{DeFelice:2015hla,DeFelice:2015moy} was built upon dRGT theory by enforcing the physical and fiducial vielbeins to be simultaneously of the Arnowitt-Deser-Misner (ADM) form and adding extra constraints to eliminate the unwanted degrees of freedom. (Minimal theory of quasidilaton massive gravity~\citep{DeFelice:2017wel,DeFelice:2017rli,DeFelice:2018vkc} is also constructed in this way upon the quasidilaton theory~\citep{DAmico:2012hia}). As the number of gravitational degrees of freedom in MTMG is only two (corresponding to tensorial gravitational waves), the theory is free from various instabilities such as the BD ghost, the Higuchi ghost, and the non-linear ghost mentioned above, and therefore provides a stable non-linear completion of the cosmological solutions in both branches of dRGT theory. In the self-accelerating branch, the graviton mass term acts as an effective cosmological constant that accelerates the expansion of the universe and the scalar perturbations behave exactly the same as in the standard $\Lambda$CDM model of GR. Moreover, the self-accelerating branch allows for GR solutions in spherical symmetry without instabilities or extra singularities~\citep{DeFelice:2018vza}. In the normal branch, on the other hand, the evolution equations for the scalar perturbations are different from those in $\Lambda$CDM and show interesting phenomenology. For example, the normal branch of MTMG may fit the redshift space distortion data better than $\Lambda$CDM~\citep{DeFelice:2016ufg} without conflicting with the integrated Sachs-Wolfe-galaxy correlation data~\citep{Bolis:2018vzs}. Furthermore, MTMG may provide a simple mechanism to enhance stochastic gravitational waves~\citep{Fujita:2018ehq,Fujita:2019tov}.

The purpose of the present paper is to explore the non-linear dynamics of MTMG for the first time by performing \mbox{$N$-body} simulations. In the self-accelerating branch, the deviation from the $\Lambda$CDM model in GR is expected to be minimal (if any). In the present paper, we therefore focus on the normal branch, which exhibits an interesting deviation from $\Lambda$CDM in the form of an environment-dependent effective gravitational constant $G_{\rm eff}$ as a function of the graviton mass and the local energy density.

The remainder of the present paper is organised as follows. In section~\ref{sec:basicequations} we briefly review MTMG and present the basic equations. After describing the implementation of the environment-dependent gravitational coupling in a modified version of the \mbox{$N$-body} code \textsc{Ramses} in Section~\ref{sec:implementation}, we calibrate the best-fit value of the MTMG parameter $\theta$ taking into account the effects of voids in section~\ref{sec:calibration}. We then explore the predictions of MTMG in the non-linear regime. In section~\ref{sec:results} we show the results of the \mbox{$N$-body} simulations in MTMG, including the power spectra, the halo mass function, the halo density profile, the halo gravitational constant profile, and the void density profile. Section~\ref{sec:summary} is then devoted to a summary of the paper and some discussions.

\section{The model and its basic equations}
\label{sec:basicequations}

The minimal theory of massive gravity (MTMG) introduced in \citep{DeFelice:2015moy} is defined through the following action:
\begin{align}
S_{\mathrm{MTMG}} & = S_{\mathrm{GR}}+\frac{\Mpl^{2}}{2}\sum_{i=1}^4\int d^4 x\, \mathcal{S}_{i}+S_{\rm constr}+S_{\rm m}\,,\label{eq:actioMTMG}\\
\mathcal{S}_{1} & = -m^{2}c_{1}\,\sqrt{\tilde\gamma}\,(N+M\mathcal{K})\,,\\
\mathcal{S}_{2} & = -\frac{1}{2}m^{2}c_{2}\,\sqrt{\tilde\gamma}(2N\mathcal{K}+M\mathcal{K}^{2}-M\tilde\gamma^{ij}\gamma_{ji})\,,\\
\mathcal{S}_{3} & = -m^{2}c_{3}\sqrt{\gamma}\,(M+N\,\mathfrak{K})\,,\\
\mathcal{S}_{4} & = -m^{2}c_{4}\sqrt{\gamma}\,N\,.
\end{align}
Here, $S_{\rm m}$ denotes a general matter action and the GR part of the action is given by the well-known ADM expression
\begin{equation}
S_{\mathrm{GR}}=\frac{\Mpl^{2}}{2}\,\int d^4xN\sqrt{\gamma}\,[{}^{(3)}R+K^{ij}K_{ij}-K^{2}]\,,
\end{equation}
where the following quantities represent the extrinsic curvature and its trace respectively:
\begin{align}
K_{ij} & = \frac{1}{2N}\,(\dot{\gamma}_{ij}-\mathcal{D}_{i}N_{j}-\mathcal{D}_{j}N_{i})\,,\\
K & = \gamma^{ij}K_{ij}\,.
\end{align}
 Here, the fields $N$ and $N^i$ are the lapse and shift, $\gamma_{ij}$ is the three-dimensional spatial metric, ${}^{(3)}R$ is the Ricci scalar of $\gamma_{ij}$, $\mathcal{D}_i$ corresponds to the covariant derivative compatible with $\gamma_{ij}$, and $\gamma\equiv\det{\gamma_{ij}}$. In the above expressions, we have also the Planck mass squared, $\Mpl^2\equiv1/(8\pi G_N)$, and $c_{1,\dots,4}$ are dimensionless constants.

Let us proceed to explaining the remaining pieces of the action written in Eq.\ (\ref{eq:actioMTMG}). To reach this goal, we need to introduce three external fields (which here we consider to be time-dependent only, for simplicity), namely $M$, $\tilde{\gamma}_{ij}$ ($\tilde\gamma$ being its determinant), and $\tilde{\zeta}^{i}{}_{j}$. The first field, $M,$ represents the fiducial lapse, the second, ${\tilde\gamma}_{ij}$, the three-dimensional fiducial metric, and the third one, $\tilde{\zeta}^{i}{}_{j}$, is related to the time-derivative of its squared root (i.e.\ the time-derivative of the fiducial vielbein). Now, out of $\gamma_{ij}$ and ${\tilde\gamma}_{ij}$, we introduce the tensor $\mathcal{K}^{m}{}_{n}$, defined by
\begin{equation}
\mathcal{K}^{m}{}_{l}\mathcal{K}^{l}{}_{n}=\tilde{\gamma}^{ms}\gamma_{sn}\,,
\end{equation}
and its inverse, $\mathfrak{K}^{m}{}_{j}$, which satisfies
\begin{equation}
\mathfrak{K}^{m}{}_{j}\mathcal{K}^{j}{}_{n}=\delta^{m}{}_{n}=\mathcal{K}^{m}{}_{j}\mathfrak{K}^{j}{}_{n}\,.
\end{equation}
We also name $\mathcal{K}\equiv\mathcal{K}^{m}{}_{m}$ and $\mathfrak{K}\equiv\mathfrak{K}^{m}{}_{m}$.

We enter the next step by introducing the constraints defined in MTMG as follows. Let us first introduce
\begin{align}
\Theta^{ij}&=\frac{\sqrt{\tilde{\gamma}}}{\sqrt{\gamma}}\{c_{1}(\gamma^{il}\mathcal{K}^{j}{}_{l}+\gamma^{jl}\mathcal{K}^{i}{}_{l})\nonumber\\
&{}+ c_{2}[\mathcal{K}(\gamma^{il}\mathcal{K}^{j}{}_{l}+\gamma^{jl}\mathcal{K}^{i}{}_{l})-2\tilde{\gamma}^{ij}]\}+2c_{3}\gamma^{ij}\,,
\end{align}
so that we can build the following scalar and vector quantities:
\begin{align}
\bar{\mathcal{C}}_{0} & = \frac{1}{2}m^{2}\,M\,K_{ij}\Theta^{ij} - m^{2}\,M\left\{\frac{\sqrt{\tilde{\gamma}}}{\sqrt{\gamma}}[c_{1}\tilde{\zeta}\right.\nonumber \\
& \left. +c_{2}(\mathcal{K}\tilde{\zeta}-\mathcal{K}^{m}{}_{n}\tilde{\zeta}^{n}{}_{m})]
+c_{3}\mathfrak{K}^{m}{}_{n}\tilde{\zeta}^{n}{}_{m}\right\},\\
\mathcal{C}^{n}{}_{i} & = -m^{2}\,M\left\{\frac{\sqrt{\tilde{\gamma}}}{\sqrt{\gamma}}\bigl[
\tfrac12 (c_1+c_2\mathcal{K})(\mathcal{K}^{n}{}_{i}+\gamma^{nm}\mathcal{K}^{l}{}_{m}\gamma_{li})\right.\nonumber\\
&-\left. c_{2}\tilde\gamma^{nl}\gamma_{li}\bigr]+c_{3}\delta^{n}{}_{i}\right\},
\end{align}
where $\tilde{\zeta}\equiv\tilde{\zeta}^{n}{}_{n}$. We are now ready to write down the last building block of the action of MTMG, namely
\begin{align}
S_{\rm constr} & = \frac{\Mpl^{2}}{2}\int d^{4}xN\sqrt{\gamma}\left(\frac{m^{2}}{4}\,\frac{M}{N}\,\lambda\right)^{\!2}\nonumber\\
&\times\left(\gamma_{ik}\gamma_{jl}-\frac{1}{2}\gamma_{ij}\gamma_{kl}\right)\Theta^{kl}\Theta^{ij}\nonumber \\
& - \frac{\Mpl^{2}}{2}\int d^{4}x\sqrt{\gamma}\left[\lambda\bar{\mathcal{C}}_{0}-(\mathcal{D}_{n}\lambda^{i})\,\mathcal{C}^{n}{}_{i}\right]\,.
\end{align}
It should be noted that the fields $\lambda$ and $\lambda^i$ are Lagrange multipliers which have been introduced as to impose four constraints. Such constraints are meant to keep the degrees of freedom of the theory on any background equal to two. This is a crucial step in the construction of MTMG.

As there is no Einstein frame for such a theory, and because it is endowed with only two gravity degrees of freedom, MTMG belongs to a type-II minimally modified gravity (MMG) theory; see also \citep{DeFelice:2020eju,Aoki:2020lig,Yao:2020tur} for other examples of type-II MMG. In other words, this construction has led to a theory which, like GR, has only two degrees of freedom. As the constraints are of scalar and vector nature, it is clear that the two tensor modes will be the propagating degrees of freedom of this theory. However, such a theory diverges from GR because of the mass of the graviton which differs from zero in general.

On an FLRW background, MTMG supports two branches, the self-accelerating branch and the normal branch. Looking for a flat FLRW solution, we give $\tilde{\gamma}_{ij}=\tilde{a}^2\,\delta_{ij}$, where $\tilde{a}$, the fiducial scale factor, represents a time-dependent external field. In this case, we can define $X\equiv\tilde{a}/a$, where $a$ is the scale factor of the physical metric, that is,\ $\gamma_{ij}=a^2\,\delta_{ij}$. Then for the self-accelerating branch, the dynamics of $X$ is bounded to satisfy $c_1X^2+2c_2X+c_3=0$. This in turn leads to a contribution to the Friedmann equation in terms of an effective cosmological constant, namely for the self-accelerating branch one has $\rho_{\rm MTMG}=\rho_{\Lambda}={\rm const}$ (this happens even when a pure cosmological constant is set to vanish from the beginning in the MTMG theory).

The second branch, called the normal branch, is the one which is considered in the present paper, and for which the following condition holds:
\begin{equation}
H=\frac{\dot{a}}{ N a} = X(t)\, \frac{\dot{\tilde{a}}}{ M \tilde{a}}\,.
\end{equation}
This leads in general to a time-dependent contribution in the Friedmann equation as follows
\begin{align}
3\Mpl^2 H^2 &= \rho_{\rm MTMG}+\sum_i\rho_i\,,\\
\rho_{\rm MTMG}(t)&=\frac{m^2\Mpl^2}2\,(3c_3X+3c_2X^2+c_1X^3)\,,
\end{align}
where $\rho_{i}$ stands for any standard matter components (including possibly a pure cosmological constant). Therefore, there are interesting possibilities as the background can acquire non-trivial deviation from GR without introducing extra degrees of freedom.

The cosmological perturbation theory of MTMG has been studied in several papers \citep{DeFelice:2016ufg,Bolis:2018vzs}, and we summarise in the following some results that will constitute the building blocks of our \mbox{$N$-body} study for MTMG.

First of all, while MTMG does not introduce any gravity degree of freedom besides the tensor modes, in the normal branch the dynamics of the matter degrees of freedom is indeed modified. In particular, if we study the dynamics of a cold and pressureless fluid, in the high-$k$ regime, and fixing the dynamics of the background to be the same as in $\Lambda$CDM\footnote{\label{footnote:MTMG-LCDMbackground}This, in the normal branch, corresponds to fixing the time-dependence of $\tilde{a}$ and $M$ so that $X=X_0={\rm const.}$ and $M=X_0\,N$ respectively. In this case, $\rho_{\rm MTMG}=\rho_{\Lambda}={\rm const.}$}  for
simplicity, we find that its energy density perturbation satisfies the following equation of motion:
\begin{equation}
\delta_m''+\left(2-\frac32\,\Omega_m\right)\delta_m'-\frac32\,\frac{G_{\mathrm{eff}}}{G_{N}}\,\Omega_m\,\delta_m=0\,,
\end{equation}
where
\begin{equation}
\frac{G_{\mathrm{eff}}}{G_{N}}=\frac{1}{1-\frac{1}{2}\theta\, Y}-\frac{1}{\left(\theta Y-2\right)^{2}}\,\frac{\rho_m}{\Mpl^2H^2}\,\theta\, Y,
\end{equation}
and $\theta$ is a free constant parameter defined as
\begin{equation}
\theta\equiv\frac{m_g^2}{H_0^2}\,,\qquad m_g^2 = \frac{X_0}2\,(c_1X_0^2+2c_2X_0+c_3)\,m^2\,.
\end{equation}
Furthermore, we have defined
\begin{equation}
Y\equiv\frac{H_{0}^{2}}{H^{2}}=\frac{3H_{0}^{2}}{8\pi G_{N}\left(\rho_{m}+\rho_{\Lambda}\right)}\,.
\end{equation}
It should be noticed that whenever $\rho_m\gg\rho_\Lambda\simeq\Mpl^2H_0^2$, then $\theta\, Y\to0$ (assuming $\theta$ to be of order unity), and $G_{\rm eff}/G_N\to1$, which corresponds to the standard Newtonian limit. This limit in cosmology corresponds to the behaviour of the perturbations in the high-redshift limit.

In the following, for the \mbox{$N$-body} simulations, we consider the previous expression for $G_{\rm eff}$ to be valid locally at every point on the three-dimensional grid. This means that the variables $Y$ and $\rho_{m}$ are to be evaluated locally in the \mbox{$N$-body} simulation. This will automatically lead to the consequence that even today, in a overdense region, we would expect $Y\ll1$. In the following we find it convenient to introduce the following dimensionless quantities:
\begin{equation}
\eta_{\Lambda}\equiv\frac{\rho_{\Lambda}}{\bar{\rho}_{m}}=\frac{\Omega_{\Lambda}\left(z\right)}{\bar{\Omega}_{m}\left(z\right)}=\frac{1-\bar{\Omega}_{m0}}{\frac{1}{a^{3}}\bar{\Omega}_{m0}}=\left(\frac{1}{\bar{\Omega}_{m0}}-1\right)a^{3}\,,
\end{equation}
which is purely time-dependent; and
\begin{equation}
\eta_{m}\equiv\frac{\rho_{m}}{\bar{\rho}_{m}}\,,
\end{equation}
so that, in overdense regions, one has $\eta_m\gg1$. Here, we consider all symbols with a bar to be evaluated on the FLRW background (or the volume average in the simulation). As a consequence, the average (background) matter density $\bar{\rho}_{m}$ is
\begin{equation}
\bar{\rho}_{m}=\frac{3\Mpl^2H_{0}^{2}\bar{\Omega}_{m0}}{a^{3}},
\end{equation}
leading to
\begin{equation}
Y=\frac{a^{3}/\bar{\Omega}_{m0}}{\eta_{m}+\eta_{\Lambda}}\,.
\label{eq:Y_ramses}
\end{equation}
Furthermore,
\begin{equation}
\frac{\rho_{m}}{3\Mpl^2 H^{2}}=\frac{\bar{\rho}_{m}}{3\Mpl^2 H^{2}}\,\eta_{m}
=\frac{H_{0}^{2}\bar{\Omega}_{m0}}{H^{2}a^{3}}\eta_{m}=Y\frac{\bar{\Omega}_{m0}}{a^{3}}\eta_{m}\,.
\end{equation}

To summarise, we implement the dynamics of the \mbox{$N$-body} simulations in \textsc{Ramses}, taking into account the effective gravitational constant, $G_{\mathrm{eff}}$, which is given locally by means of the following relation:
\begin{equation}\label{eq:Geff1}
\frac{G_{\mathrm{eff}}}{G_{N}}=\frac{1}{1-\frac{1}{2}\theta Y}-\frac{3\bar{\Omega}_{m0}\theta Y^{2}\eta_{m}}{a^{3}\left(\theta Y-2\right)^{2}}\,,
\end{equation}
with $Y$ being given by Eq.\ (\ref{eq:Y_ramses}). Now we are ready to implement MTMG in our \mbox{$N$-body} simulations, so that we can start exploring the behaviour of gravity in MTMG in non-linear regimes.

\section{Massive gravity implementation in \textsc{Ramses}}
\label{sec:implementation}

To quantify the effects of MTMG in the cosmological evolution of structures, we run a set of cosmological simulations. The simulations are performed with a modified version of the \mbox{$N$-body} code \textsc{Ramses} (Teyssier 2002).

The standard GR version of \textsc{Ramses} solves Poisson's equation $\nabla^{2}\Phi=4\pi G\delta\rho$ to find $\Phi$ at the centre of each grid cell. We note that $\delta\rho=\rho-\bar{\rho}$ can be either positive (for overdensities) or negative (for underdense regions). Solving the Poisson equation for a given distribution of matter gives us a continuous $\Phi$ field. One can then calculate the acceleration at the position of each particle as $\ddot{\mathbf{x}}=-\nabla\Phi$. The value of $\nabla\Phi$ at the particle position is guessed by CIC linear interpolation from nearby grid cell centres (where the value is known). The particles are then moved one step in time using forward time integration with position, velocity, and acceleration.

In the case of MTMG, the Poisson equation is modified to $\nabla^{2}\Phi=4\pi G_{\mathrm{eff}}\left(\rho\right)\delta\rho$. One  then needs to transform $G\rightarrow G_{\mathrm{eff}}$. This can be done either directly in the Poisson's equation or when calculating the acceleration of each particle. We use the local $G_{\mathrm{eff}}$ in the Poisson equation when calculating the gravitational potential field. In this way, local gradients of the potential field, and hence the acceleration of matter, will be modified by structures elsewhere. This gives the correct long-range forces, where two clusters separated by a void feel the $G_\mathrm{eff}$ encoded in the void (see section~\ref{sec:calibration}).

The initial matter distribution is generated with the package Grafic (Bertschinger 2001) with standard gravity. The approximation that we make without including modified gravity in the initial conditions is justified by the fact that modifications to GR occur only at much lower redshifts. All simulations use the same initial matter distribution and assume a flat $\Lambda$CDM background cosmology (see footnote~\ref{footnote:MTMG-LCDMbackground} for MTMG in this respect) provided by the Planck collaboration: $\Omega_m = 0.3175$, $\Omega_\Lambda = 0.6825$, and $H_0 = 67.11$ km/s/Mpc (Planck Collaboration et al.\ 2018). The number of particles is $256^3$, and the size of the box is 64 Mpc/h.

\section{Calibration of graviton mass}
\label{sec:calibration}

From a phenomenological viewpoint, one of the most important aspects of MTMG in the normal branch is that the effective gravitational constant $G_{\rm eff}$ depends on the environment. In linear perturbation theory, $G_{\rm eff}$ is a function of the energy density of the background FLRW universe and thus is homogeneous in space. As explained in the previous sections, in order to explore the non-linear dynamics of the normal branch of MTMG we promote $G_{\rm eff}$ to a function of a coarse-grained energy density so that it depends not only on the time but also on the spatial position and the coarse-graining scale. As $G_{\rm eff}$ is a non-linear function of the energy density, the spatial average of $G_{\rm eff}$ does not agree with $G_{\rm eff}$ for the averaged energy density, that is, $\langle G_{\rm eff}(m_g^2, \rho) \rangle \ne G_{\rm eff}(m_g^2, \langle \rho\rangle)$, where $m_g^2$ is the graviton mass squared, $\rho$ is the energy density, $\langle X \rangle$ represents the volume average of a local function $X$. This makes it non-trivial to compare results from the non-linear simulation with predictions of the linear perturbation theory even at the largest scales.

To understand this point, suppose that there are two groups of particles separated from each other by a void region and that we would like to compute the gravitational force between a particle in one group and another particle in the other group. If the spatial size of each group is sufficiently small compared with the separation between the two groups then the strength of the gravitational force should be computed using the value of $G_{\rm eff}$ in the void region that separates the two groups. This means that the non-linear dynamics at largest scales should reflect the value of $G_{\rm eff}(m_g^2, \rho_{\rm void}(z))$, where $\rho_{\rm void}(z)$ is the typical energy density in void regions at the redshift $z$. On the other hand, the predictions of the linear perturbation theory reflect the value of $G_{\rm eff}(m_g^2, \rho_{\rm FLRW}(z))$, where $\rho_{\rm FLRW}(z) = \langle \rho_{\rm local}(z,\vec{x})\rangle$ is the volume-averaged energy density, which corresponds to the FLRW background density at the redshift $z$. Here, $\rho_{\rm local}(z,\vec{x})$ is the local density at the redshift $z$ and the spatial position $\vec{x}$. In particular, the best-fit value of the graviton mass squared $m_g^2$ was obtained using $G_{\rm eff}(m_g^2, \rho_{\rm FLRW}(z))$. Therefore, provided that non-linear voids develop sufficiently at the redshift $z=z_{\rm obs}$ relevant for the observational bounds on $m_g^2$, we need to calibrate $m_g^2$ as $m_g^2$ as $G_{\rm eff}(m_{\rm nl}^2, \rho_{\rm void}(z_{\rm obs})) = G_{\rm eff}(m_{\rm lin}^2, \rho_{\rm FLRW}(z_{\rm obs}))$, where $m_{\rm lin}^2$ is the best-fit value of $m_g^2$ that was obtained using the prediction of the linear theory, and $m_{\rm nl}^2$ is the calibrated graviton mass squared for the non-linear dynamics. In practice, we implement this idea of calibration as
\begin{equation}
G_{\rm eff,avg}(m_{\rm nl}^2, z_{\rm obs}) = G_{\rm eff}(m_{\rm lin}^2, \rho_{\rm FLRW}(z_{\rm obs}))\,,
\end{equation}
where
\begin{equation}
G_{\rm eff,avg}(m_g^2, z) \equiv
\langle G_{\rm eff}(m_g^2, \rho_{\rm local}(z,\vec{x}))\rangle
\end{equation}
is the volume-averaged effective gravitational constant. For $m_{\rm lin}^2 \simeq -3.828 H_0^2$ obtained from the RSD data and the integrated Sachs-Wolfe-galaxy correlation data~\citep{DeFelice:2016ufg,Bolis:2018vzs}, $z_{\rm obs} \simeq 0$ (to be more precise, $z_{\rm obs} \simeq 0{-}0.6$ but we set $z_{\rm obs} \simeq 0$ for simplicity) and $G_{\rm eff}(m_{\rm lin}^2, \rho_{\rm FLRW}(z_{\rm obs})) \simeq 0.45 \times G_{\rm N}$ (see Fig.~4 of \citep{DeFelice:2016ufg}), and thus
\begin{equation}
m_{\rm nl}^2 \simeq -1.7 H_0^2\,.
\end{equation}
Only after this calibration of the graviton mass can we make predictions by \mbox{$N$-body} simulations. Therefore, in the rest of the present paper, we adopt this value.

\subsection{Average $G_\mathrm{eff}$ as a function of time}

\begin{figure}[!htb]
\includegraphics[width=0.9\columnwidth]{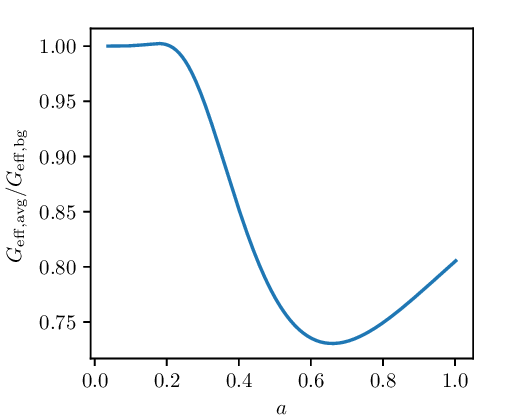}
\caption{Volume averaged $G_\mathrm{eff}$ compared to the $G_\mathrm{eff}$ expected from the background $\rho_m$. The scale factor $a$ is along the horizontal axis. Here, $\theta = -1.7$. At early times when voids have not yet developed, effects of voids are negligible and therefore the averaged $G_\mathrm{eff}$ and $G_\mathrm{eff}$ for the average density agree with each other. On the other hand, at late times when voids are present, they deviate from each other due to the lower density in voids.
\label{fig:Geff_converg}}
\end{figure}

In figure \ref{fig:Geff_converg} we plot the volume-averaged $G_\mathrm{eff}$ compared to the $G_\mathrm{eff}$ expected from the background $\rho_m$ as a function of the scale factor. Here, we assume as an example $\theta = -1.7$ because this is the calibrated best-fit value as explained in the previous paragraph. The necessity of the calibration, namely the difference between the volume-averaged $G_\mathrm{eff}$ and the $G_\mathrm{eff}$ for the averaged density, stems from the existence of voids where the density is lower than the average. At early times, voids have not yet developed and therefore the volume-averaged $G_\mathrm{eff}$ and the $G_\mathrm{eff}$ for the averaged density agree with each other. On the other hand, at late times, as voids develop, the volume-averaged $G_\mathrm{eff}$ begins to deviate from the $G_\mathrm{eff}$ for the averaged density. After a slight increase, the former starts to decrease significantly compared with the latter, as expected from Fig.~4 of \cite{DeFelice:2016ufg}, and as clearly seen in figure \ref{fig:Geff_converg} of this paper. The $\mathcal{O}(1)$ difference between the two quantities at $a=1$ clearly shows that the calibration is necessary to match predictions of the linear perturbation theory and those of non-linear \mbox{$N$-body} simulations.

\subsection{Void gravitational constant profile}

\begin{figure}[!htb]
\includegraphics[width=1\columnwidth]{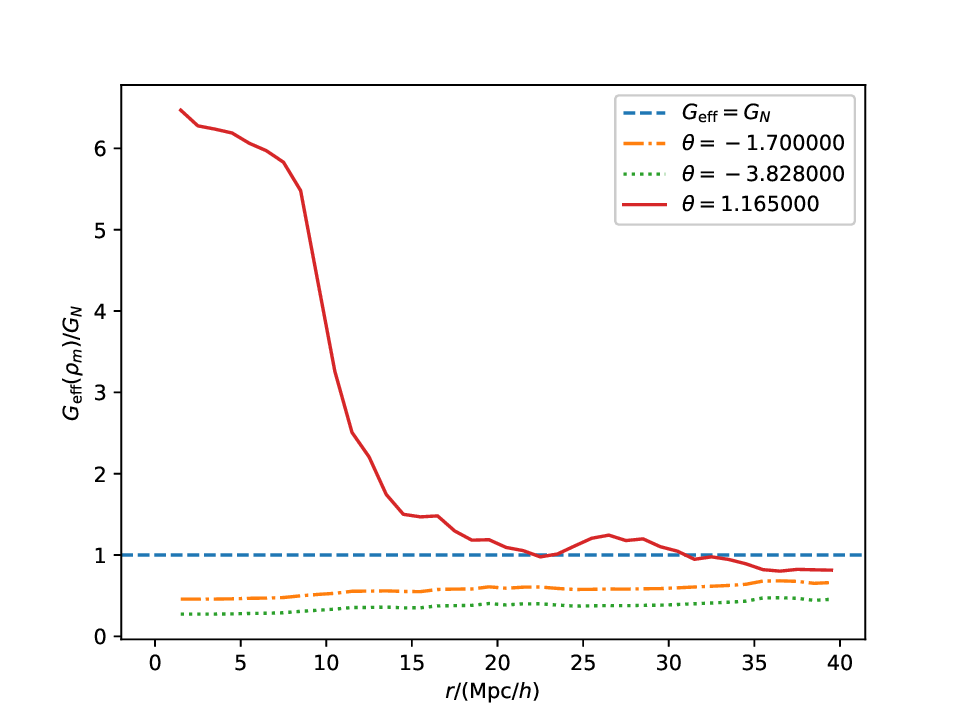}
\caption{Profile of $G_\mathrm{eff}$ around the centre of a deep void for $\theta = -1.7$, $\theta = -3.828$ and $\theta = 1.165$. For negative values of $\theta$, the effective gravitational constant, $G_\mathrm{eff}$, is always smaller than $G_\mathrm{N}$ around voids. \label{fig:geff_r_void}}
\end{figure}

In figure \ref{fig:geff_r_void} we plot the radial profile of $G_\mathrm{eff}$ around the centre of a deep void for $\theta = -1.7$, $\theta = -3.828,$ and $\theta = 1.165$. For negative $\theta,$ the effective gravitational constant, $G_\mathrm{eff}$, is always smaller than $G_\mathrm{N}$ in low-density regions (such as voids). While for positive $\theta,$ the effective gravitational constant, $G_\mathrm{eff}$, is larger than $G_\mathrm{N}$ in low-density regions. Models with a positive $\theta$ (resulting in increased $G_\mathrm{eff}$ in voids) are ruled out by the integrated Sachs-Wolfe-galaxy correlation data~\citep{Bolis:2018vzs} and are not be considered any further in this work.

\section{Results}
\label{sec:results}

We performed three different simulations: GR ($\theta=0$), MTMG with $\theta=-3.828$, and MTMG with $\theta=-1.7$.
For these choices of parameters, we ran high-resolution simulations with $1024^{3}$ particles and a box of 256 Mpc/$h$. Halos were calculated with the Rockstar halo finder \citep{Behroozi:2013ApJ}.

\subsection{Power spectra}

\begin{figure}[!htb]
\includegraphics[width=0.9\columnwidth]{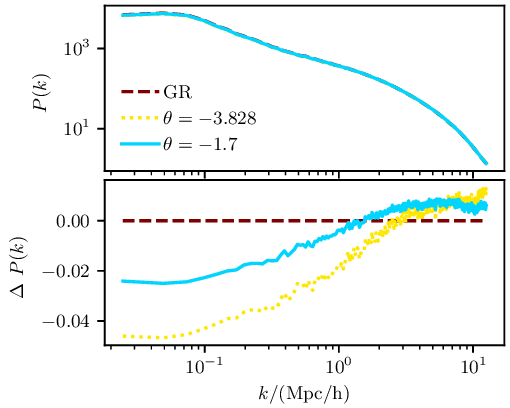}
\caption{ Small-scale power spectra at $z=0$ for the high-resolution simulations. A suppression in the power spectra in the quasi-non-linear scales is present. As one enters the fully non-linear scales the power spectra approaches GR. \label{fig:Pk_z0_hr}}
\end{figure}

In figure \ref{fig:Pk_z0_hr} we plot the matter power spectrum for two different negative values of $\theta$ at redshift $z=0$. The large-scale linear regime was studied in previous works~\citep{DeFelice:2016ufg,Bolis:2018vzs} and the constraints obtained by the linear perturbation theory are valid as long as the graviton mass, or the parameter $\theta$, is calibrated properly as explained in section~\ref{sec:calibration}. For this reason, we focus on the non-linear small scales. From the figure, one can see that there is a suppression in the power spectra in the quasi-non-linear scales. As one enters the fully non-linear scales the power spectra approach GR, as we expect GR to be fully recovered at the very small-scale and high-density regions (in accordance with Fig.~4 of \citet{DeFelice:2016ufg}) when one approaches early times and high densities.

\subsection{Halo mass function, mass histogram}

\begin{figure}[!htb]
\includegraphics[width=0.9\columnwidth]{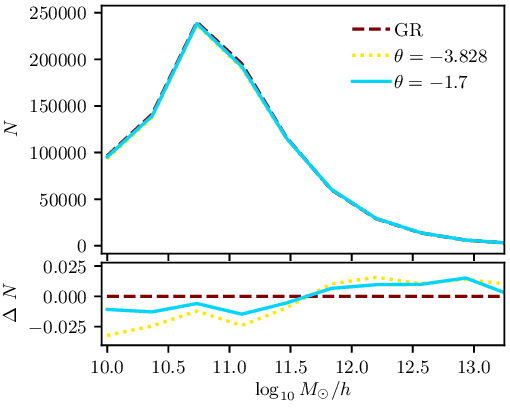}
\caption{Halo mass function. Lower panel is relative difference with respect to the same population in GR. There are more massive halos in MTMG, while there is a suppression of the abundance of small halos.}
\label{halomass}
\end{figure}

Figure \ref{halomass} shows the halo mass function for MTMG. The lower panel shows the relative difference with respect to the same population in GR. One can see that MTMG creates a larger amount of massive halos in general, while there is a suppression of the abundance of small halos. This can be understood by taking into account the fact that smaller halos reside in relatively dense regions, where the effective gravitational constant can be larger than $G_N$, i.e.\ $G_{\mathrm{eff}} \gtrsim G_{N}$ (although $G_{\mathrm{eff}} \to G_{N}$ in the high-density limit; see figure \ref{fig:geff_r_ld}), and so the merger rate for substructures is higher than GR; small halos will interact to form larger halos at a higher rate.

\subsection{Halo density profile}

\begin{figure}[!htb]
\includegraphics[width=0.9\columnwidth]{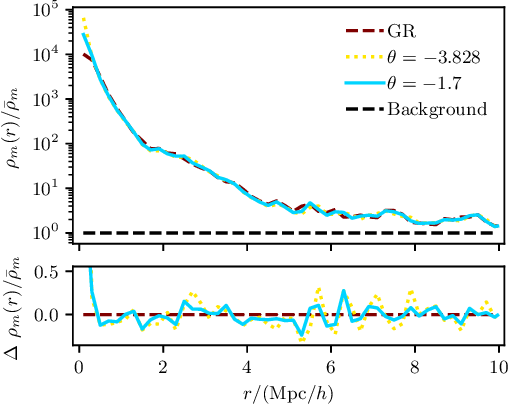}
\caption{ Dark matter halo density profile for GR and different negative $\theta$. The black dashed line is the average background density. The lower panel is the difference with respect to GR. \label{fig:rho_r_ld}}
\end{figure}

Figure \ref{fig:rho_r_ld} shows the dark matter density profile in a halo. The density is calculated as an average within concentric shells of some thickness $\Delta r$, centred on the coordinates of a given halo in the GR simulation. We expect similar halos to form at approximately the same coordinate in the other simulations because the initial particle distributions are identical. The halos have virial masses of $10^{14} - 10^{15} \, M_\odot/h$ and virial radii of about $2.0 \, \mathrm{Mpc}/h$. It is clear from this figure that the density profile is not a good probe for MTMG because the differences are small. We note that as one approaches the centre of the halo the density profile becomes similar to the GR one. This is expected as in this theory $G_{\mathrm{eff}} \sim G_{N}$ in very high-density environments. On the other hand, at the outskirts of the halo, where the density of halos are lower and close to the critical density, $G_{\mathrm{eff}} < G_{N}$ and so the effects on the dark matter halo profiles differ from GR.

\subsection{Halo gravitational constant profile}

\begin{figure}[!htb]
\includegraphics[width=0.9\columnwidth]{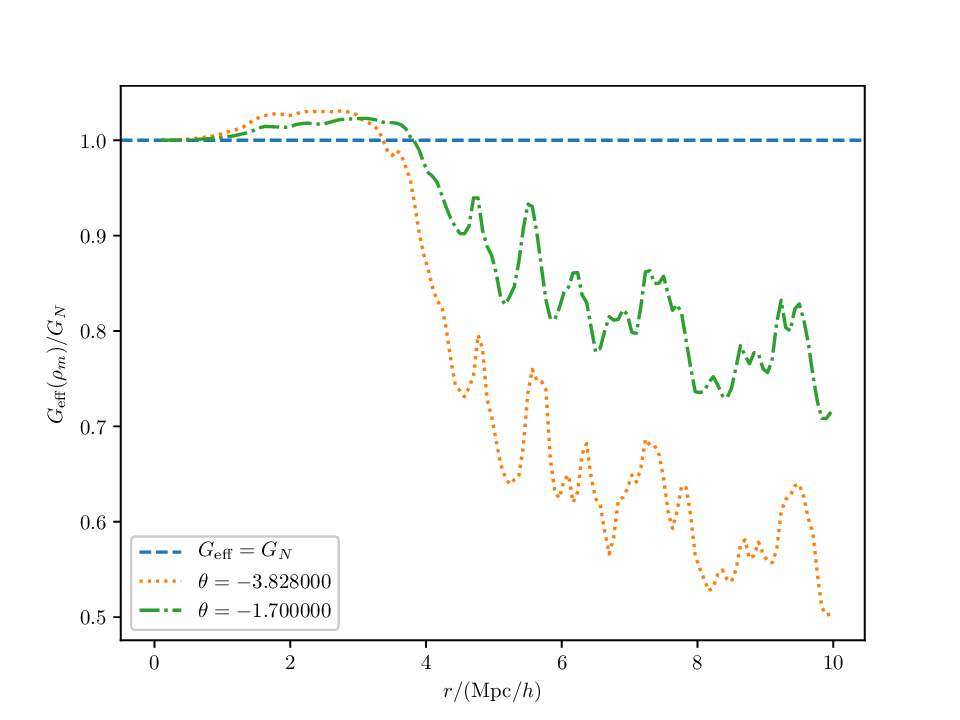}
\caption{Profile of $G_\mathrm{eff}$ around an isolated halo. Towards the centre of the halo, where the density is high, $G_\mathrm{eff}$ approaches $G_N$. On the other hand, in the outskirts of the halo $G_\mathrm{eff}$ becomes smaller than $G_N$. This effect is stronger for more negative values of $\theta$. \label{fig:geff_r_ld}}
\end{figure}

In figure \ref{fig:geff_r_ld} we show the effective gravitational constant $G_\mathrm{eff}$, which is found by applying equation \eqref{eq:Geff1} to the density profiles around a halo.
We see that $G_\mathrm{eff}$ tends towards $G_N$ in the highest density regions, signifying a working screening. For intermediate densities in the outskirts of the halo (around $2-3 \, \mathrm{Mpc}/h$ or $1-2\,r_\mathrm{vir}$ from the centre), we see a slight enhancement of $G_\mathrm{eff}$ over $G_N$, which can lead to increased clustering at intermediate to small scales. As $r$ further increases, $G_{\rm eff}/G_N$ tends to decrease and to have some oscillations. This behaviour is due to a decrement of the local environmental energy density (as moving out from the centre of the halo) and the peaks of the oscillations correspond to the presence of subhalos whose local density has some local peak.

\subsection{Void density profile}

\begin{figure}[!htb]
\includegraphics[width=0.9\columnwidth]{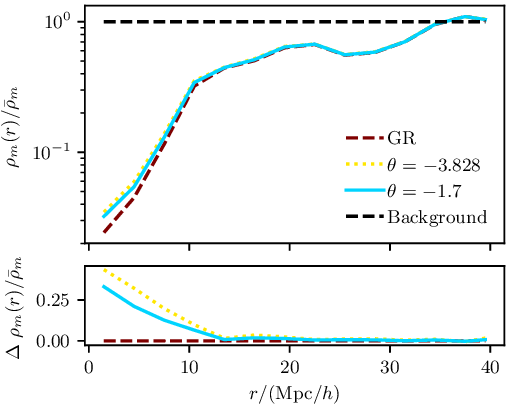}
\caption{Density profile centred at a deep void. Voids are denser in MTMG than in GR, and their density profile depends on the value of $\theta$. \label{fig:rho_r_void}}
\end{figure}

Figure \ref{fig:rho_r_void} shows the density profile of a void. The density is calculated as an average within concentric shells of some thickness $\Delta r$, centred on the coordinates of a given void in the GR simulation. We expect similar voids to form at approximately the same coordinate in the other simulations because the initial particle distributions are identical. A clear difference between MTMG and GR is that voids are denser near their centres in MTMG than in GR. This is because in MTMG, $G_\mathrm{eff}$ is lower than $G_N$ in voids and so particles deep inside voids will not feel a force that attracts them to the boundaries of the voids (where the density is higher) as strongly as in GR.

\section{Summary and discussion}
\label{sec:summary}

We performed {\mbox{$N$-body}} simulations in the normal branch of the minimal theory of massive gravity (MTMG), employing the \textsc{Ramses} code and implementing an environment-dependent effective gravitational constant $G_{\rm eff}$ as a function of the graviton mass and the local energy density as predicted in MTMG. We show how the effective gravitational coupling $G_\mathrm{eff}$ changes within voids and dark matter halos.

We find that halo density profiles are not a good probe for MTMG because deviations from GR are small. This is of no surprise because we expect MTMG to be screened, and to recover GR in high-density regions. Similarly, the matter power spectra show deviations only at the percentage level.

A clear difference between MTMG and GR is that voids are denser in MTMG than in GR. This is because in MTMG, $G_\mathrm{eff}$ is lower than $G_N$ in voids, such that particles deep inside voids will not feel a strong force that attracts them to the structures outside of the voids (where the density is higher). As measuring voids profiles is currently a relatively complex task from an observational point of view, a better probe of MTMG would be the halo abundances. We find that MTMG creates a larger amount of massive halos, while there is a suppression of the abundance of small halos. This phenomenon is due to the fact that in overdense regions, in MTMG, we have $G_{\rm eff}\gtrsim G_N$ (and $G_{\rm eff}\to G_N$ in very high-density regions). This leads, in general, to a stronger gravitational interaction among small subhalos in overdense regions, which in turn tends to build up a larger number of massive halos through mergers compared to GR.

The MTMG model with an environmentally dependent $G_\mathrm{eff}$ provides a framework rich in phenomenology. In this work, we find that MTMG has signatures distinguishable from GR on non-linear scales, while still being a good fit to current observations. Future observations and further studies of massive gravity in the non-linear regime will provide further insights into the nature of gravity.

\begin{acknowledgements}
RH and DFM thank the Research Council of Norway for their support. Computations were performed on resources provided by UNINETT Sigma2 -- the National Infrastructure for High-Performance Computing and Data Storage in Norway. The work of ADF was supported by Japan Society for the Promotion of Science Grants-in-Aid for Scientific Research No.~20K03969. The work of SM was supported by Japan Society for the Promotion of Science Grants-in-Aid for Scientific Research No.~17H02890, No.~17H06359, and by World Premier International Research Center Initiative, MEXT, Japan.
\end{acknowledgements}

\end{document}